  \documentclass[pdftex]{llncs}
\usepackage{amsmath}
\usepackage{amssymb}
\usepackage{llncsdoc}
\usepackage{ifpdf}

\usepackage{subfigure}
\usepackage{float}
\usepackage{makeidx}

\ifpdf
  \usepackage[pdftex]{epsfig}
  \usepackage[%colorlinks=true,
              citecolor=blue,
              urlcolor=blue,
              bookmarksopen=true,
              bookmarksnumbered=true,
              raiselinks=true,
              pdfstartview=FitH]{hyperref}
\else
  \usepackage{epsfig}
\fi

\date{}

\newcommand{\marin}{\textsc{marin}}

\begin{document}

\mainmatter              % start of the contributions

\title{Experiments on extreme wave generation  \\
       using the Soliton on Finite Background}
\author{Ren\'{e} H.M.\  Huijsmans\inst{1}
  \and Gert Klopman\inst{2,3}
  \and Natanael Karjanto\inst{3}
  \and Andonawati\inst{4} }
\authorrunning{Huijsmans, Klopman, Karjanto and Andonowati}
\institute{Maritime Research Institute Netherlands, Wageningen,
           The Netherlands, \\
           e-mail: \email{r.h.m.huijsmans@marin.nl}
\and Albatros Flow Research, Marknesse, The Netherlands
\and Applied Analysis and Math.\  Physics, Univ.\  of Twente, Enschede,
     The~Netherlands
\and Jurusan Matematika, Institut Teknologi Bandung, Bandung, Indonesia}
\maketitle

%\pagestyle{empty}
%%%%%%%%%%%%%%%%%%%%%%%%%%%%%%%%%%%%%%%%%%%%%%%%%%%%%%%%%%%%%%%%%%%%%%%%%%%%%%%%%%%%%%%%%%%%%%%%%%%%%%%%%%%%%%%%%%%%%%%%%%%%%%%%%%%%%%%%%%%%%%%%%%%%%%%%%
%%%%%%%%%%%%%%%%%%%%%%%%%%%%%%%%%%%%%%%%%%%%%% To be written for a tiny fraction of Rogue Waves 2004 paper %%%%%%%%%%%%%%%%%%%%%%%%%%%%%%%%%%%%%%%%%%%%%%
%%%%%%%%%%%%%%%%%%%%%%%%%%%%%%%%%%%%%%%%%%%%%%%%%%%%%%%%%%%%%%%%%%%%%%%%%%%%%%%%%%%%%%%%%%%%%%%%%%%%%%%%%%%%%%%%%%%%%%%%%%%%%%%%%%%%%%%%%%%%%%%%%%%%%%%%%

%\maketitle
\section*{Introduction}

Freak waves are very large water waves whose heights exceed the
significant wave height of a measured wave train by a factor of more
than 2.2. However, this in itself is not a well established definition
of a freak wave. The mechanism of freak wave generation in reality as
well as modeling it in a wave basin has become an issue of great
importance.\\
Recently one is aware of the generation of freak wave through the
Benjamin--Feir type of instability or self--focussing. Consequently the
Non--Linear--Schr\"{o}dinger (NLS) equation forms a good basis for
understanding the formation of freak waves. However, the complex
generation of a freak wave in nature within a sea condition is
still not well understood, when the non-linearity of the carrier wave is
not small.  In our study we will focus on the Soliton on Finite
Background, an exact solution of the NLS equation, as a generating mechanism
for extreme waves.\\
Apart from a numerical investigation into the evolution of a soliton on
a finite background also extensive detailed model tests have been performed
for validation purposes in the hydrodynamic laboratories of the Maritime
Research Institute Netherlands (\marin).
Furthermore, a numerical wave tank \cite{West01} is used to model
the complete non-linear non-breaking wave evolution in the basin.

\section*{Properties of the Soliton on Finite Background}

The NLS equation is chosen as a mathematical model for the non-linear
evolution of the envelope of surface wave packets. For spatial evolution
problems, it is given in non-dimensional form and in a frame of
reference moving with the group velocity by
\begin{equation}
  \partial_{\xi}\psi + i\beta\partial_{\tau}^{2}\psi + i\gamma|\psi|^{2}\psi = 0,
\end{equation}
where $\xi$ and $\tau$ are the corresponding spatial and temporal
variables, respectively; $\beta$ and $\gamma$ are the dispersion
and non-linearity coefficients. This equation has
many families of exact solutions. One family of exact
solutions is known as the Soliton on Finite Background (SFB) and this
is a good candidate for describing extreme waves. This exact
solution has been found by Akhmediev, Eleonski$\breve{\textmd{i}}$ \& Kulagin
\cite{AkEl87}, see also \cite{AkKo86} and \cite{AkAn97}.

This SFB solution describes the dynamic evolution of an unstable
modulation process, with dimensionless modulation frequency $\nu$.  In
the context of water waves, for infinitesimal modulational
perturbations to a finite-amplitude plane wave,  this process is known
as Benjamin-Feir (BF) instability \cite{BeFe67}.
%This phenomenon says that
%under a very long modulation with a certain condition, the plane
%wave (or the continuous wave) solution of the NLS equation will
%grow exponentially.
However, non-linearity will limit this exponential growth and the SFB
is one (of many other) non-linear extension of the BF instability for
larger amplitudes of the modulation.
Extensive research on the NLS equation and the SFB solution, to obtain
a better understanding of deterministic extreme-wave phenomena has been
conducted in the past few years (see \textit{e.g.} \cite{OsOS00},
\cite{Osbo01}, \cite{AnKG04} and \cite{GrAK04}).

An explicit expression for the SFB is given as the following
complex-valued function
\begin{equation}
  \psi(\xi,\tau;\tilde{\nu},r_{0}) = A(\xi) \cdot
  \left\{
    \frac{ \tilde{\nu}^{2}\cosh (\sigma \xi)
           - i \left[ \sigma/(\gamma r_{0}^{2})\right]\;\sinh (\sigma \xi)}
         { \cosh (\sigma \xi)
           - \sqrt{1 - \frac{1}{2}\tilde{\nu}^{2}} \cos(\nu \tau)}
    - 1
  \right\}, \label{2}
\end{equation}
where $A(\xi) = r_{0}e^{-i\gamma r_{0}^{2}\xi}$ is the
\textit{plane-wave} or the \textit{continuous wave} solution of
the NLS equation, $\sigma = \gamma
r_{0}^{2}\tilde{\nu}\,\sqrt{2-\tilde{\nu}^2}$ is the
\textit{growth rate} corresponds to the Benjamin-Feir instability,
$\nu = \tilde{\nu}\,r_{0}\sqrt{\frac{\gamma}{\beta}}$ is the
modulation frequency, and $\tilde{\nu}$, $0 < \tilde{\nu} <
\sqrt{2}$ is the normalized modulation frequency. This SFB reaches
its maxima at $(\xi,\tau) = \left(0, \frac{2n\pi}{\nu}\right)$, $n
\in \mathbb{Z}$. It has a soliton-like form with a finite
background in the spatial $\xi$-direction. The SFB is periodic along in
the temporal $\tau$-direction, with period
$\frac{2\pi}{\nu}$. For $|\xi| \rightarrow \infty$, the SFB turns
into the continuous wave solution $A(\xi)$. It possesses two
essential parameters: $r_{0}$ and $\tilde{\nu}$.
%Another
%parameter, the carrier wave frequency $\omega_{0}$, takes into
%account in the corresponding physical wave packet profile.
\begin{figure}[ht]
  \begin{center}
    \ifpdf
      \epsfig{file=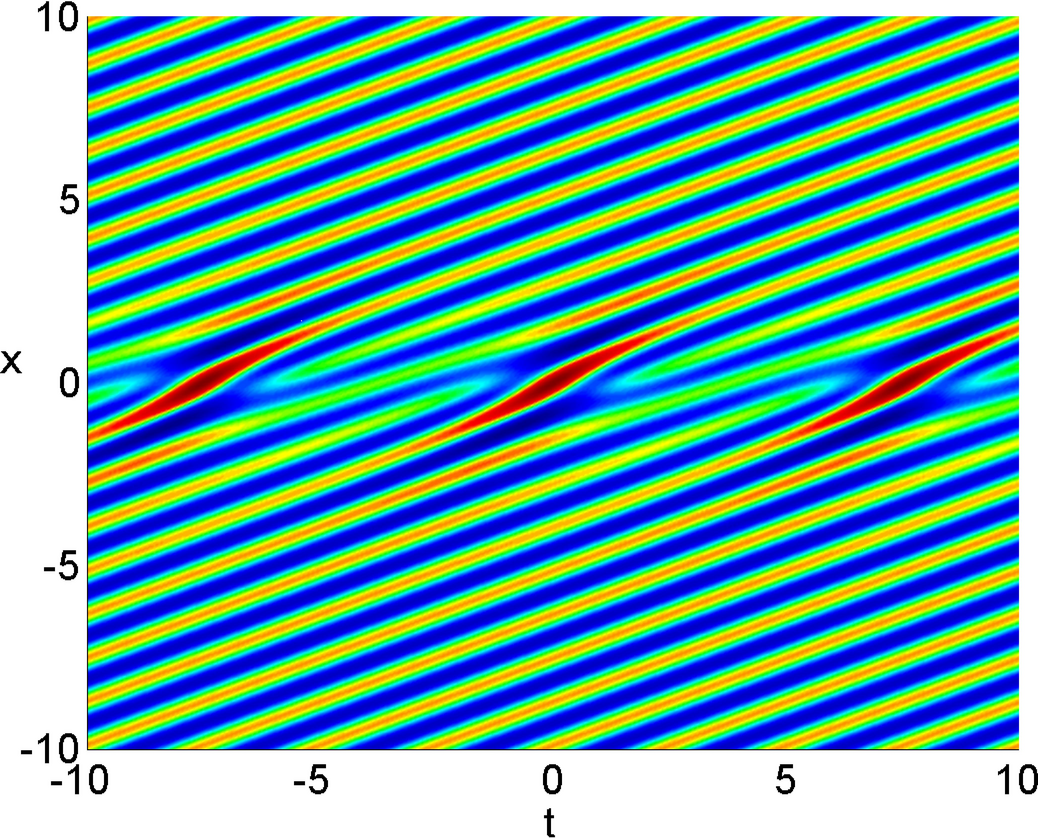,width=0.40\textwidth}
    \else
      \epsfig{file=SFB_physical-rhmh.eps,width=0.40\textwidth}
    \fi
    \caption{\footnotesize{Density plots of a physical wave packet profile
             according to an SFB envelope
             for $\tilde{\nu}_{1} = 1$, showing the wave dislocation
             phenomenon.}}
    \label{SFBphysical}
  \end{center}
\end{figure}

The first-order part of the corresponding physical wave packet profile
$\eta(x,t)$ for a given complex-valued function $\psi(\xi,\tau)$ is expressed
as follows
\begin{equation}
  \eta(x,t) = \psi(\xi,\tau) e^{i(k_{0}x - \omega_{0}t)} +
  \textrm{c.c.},
\end{equation}
where c.c. means the complex conjugate of the preceding term, the
wave number $k_{0}$ and frequency $\omega_{0}$ satisfy the linear
dispersion relation $\omega = \Omega(k) \equiv \sqrt{k\tanh k}$.
The variables $(x,t)$ in the non-moving frame of reference are related
to $(\xi,\tau)$ in the moving frame of reference by the transformation
$\xi = x$ and $\tau = t - x/\Omega'(k_{0})$.
The modulus of $\psi$ represents the wave
group envelope, enclosing the wave packet profile $\eta(x,t)$. The
dimensional laboratory quantities are related to the non-dimensional
quantities by the following Froude scaling, using the gravitional acceleration
$g$ and the depth of the basin $h$: $x_{\textmd{lab}} = x \cdot h$,
$t_{\textmd{lab}} = t \cdot \sqrt{\frac{g}{h}}$, $k_{\textmd{lab}}
= k/h$, $\omega_{\textmd{lab}} = \omega \cdot \sqrt{\frac{g}{h}}$,
and $\eta_{\textmd{lab}} = \eta \cdot h$.

In principle, the wave profile including the higher-order terms
represents a good approximation to the situation in real life. To
accommodate this fact, we will include higher-order terms up to second order.
We apply an perturbation-series
expansion (Stokes' expansion) to the physical wave-packet profile
$\eta(x,t)$ and the multiple-scale approach using the variables $\xi$
and $\tau$, where $\xi = \epsilon^{2} x$, $\tau = \epsilon(t -
x/\Omega'(k_{0}))$, and $\epsilon$ is a small positive non-linearity and
modulation parameter.
The corresponding physical wave packet profile, consisting of the
superposition of the first-order harmonic term of
\cal{O}$(\epsilon)$ and a second-order non-harmonic long wave as
well as a second-order double-frequency harmonic term of
\cal{O}$(\epsilon^{2})$, is given by
\begin{eqnarray}
  \eta(x,t) &=& \epsilon \left[
    \psi^{(10)}(\xi,\tau) e^{i(k_{0}x - \omega_{0}t)} + \textrm{c.c.}
  \right]
  \nonumber \\
  &+& \epsilon^{2}\left\{
      \psi^{(20)}(\xi,\tau)
      + \left[
        \psi^{(22)}(\xi,\tau) e^{2i(k_{0}x - \omega_{0}t)}
         + \textrm{c.c.}
      \right]
    \right\}.
\end{eqnarray}
%By substituting the surface wave elevation $\eta(x,t)$ into the KdV
%equation with exact linear dispersion relation, $\partial_{t}\eta
%+ i \Omega(-i\partial_{x})\eta + \frac{3}{4}\partial_{x}\eta^{2} =
%0$, as a governing equation for $\eta$, then we find that
We find from the multiple-scales perturbation-series approach that
$\psi^{(10)}(\xi,\tau)=\psi(\xi,\tau)$ satisfies the spatial NLS equation and
\begin{eqnarray}
  \psi^{(20)}(\xi,\tau) &=& - \frac{1}{\Omega(k_0)}\;
      \frac{4 k_0 \Omega'(k_0) - \Omega(k_0)}
           {[\Omega'(0)]^2 - [\Omega'(k_0)]^2}\;
      |\psi^{(10)}(\xi,\tau)|^{2}\\
  \psi^{(22)}(\xi,\tau) &=&
    k_0\; \frac{ 3 - \tanh^2 k_0 }{2 \tanh^3 k_0}\;
    [\psi^{(10)}(\xi,\tau)]^{2}.
\end{eqnarray}
%\begin{eqnarray}
%  \psi^{(20)}(\xi,\tau) &=& \frac{3}{4}       \frac{|\psi^{(10)}(\xi,\tau)|^{2}}{\Omega'(k_{0}) - \Omega'(0)}, \\
%  \psi^{(22)}(\xi,\tau) &=& \frac{3}{2} k_{0} \frac{ \psi^{(10)}(\xi,\tau) ^{2}}{2\omega_{0} - \Omega(2k_{0})}.
%\end{eqnarray}
A similar derivation for the temporal NLS equation resulting from the
KdV equation can be found
in \cite{Groe98}. By including this second-order term, the wave signal
$\eta(x,t)$ experiences the well-known Stokes' effect: the crests become
steeper and the troughs becomes shallower \cite{Debn94}.

%%% NEW! Added on October 21, 2011.
The coefficients $\beta$ and $\gamma$ of the spatial NLS equation are given, in nondimensional form, as:
\begin{eqnarray}
  \beta  &=& -\frac{1}{2} \frac{\Omega''(k_0)}{[\Omega'(k_0)]^3}, \\
  \gamma &=& \frac{\gamma_1 + k_0 \alpha_U + \lambda \alpha_\zeta}{\Omega'(k_0)},
\end{eqnarray}
where
\begin{eqnarray}
  \gamma_1 &=& k_0^2 \Omega(k_0) \frac{9 \tanh^4 k_0 - 10 \tanh^2 k_0 + 9}{4 \tanh^4 k_0}, \\
  \lambda  &=& \frac{1}{2} k_0^2 \frac{1 - \tanh^2 k_0}{\Omega(k_0)}, \\
  \alpha_\zeta &=& -\frac{1}{\Omega(k_0)} \frac{4k_0 \Omega'(k_0) - \Omega(k_0)}{[\Omega'(0)]^2 - [\Omega'(k_0)]^2} \qquad \textmd{and} \\
  \alpha_U &=& \alpha_\zeta \Omega'(k_0) - \frac{2k_0}{\Omega(k_0)}.
\end{eqnarray}
These can be used to compute the SFB solution $\psi(\xi,\tau)$ from Equation~\eqref{2}.

\subsection*{Phase singularity and wave dislocation}
\begin{figure}[ht]
  \begin{center}
    \ifpdf
      \epsfig{file = 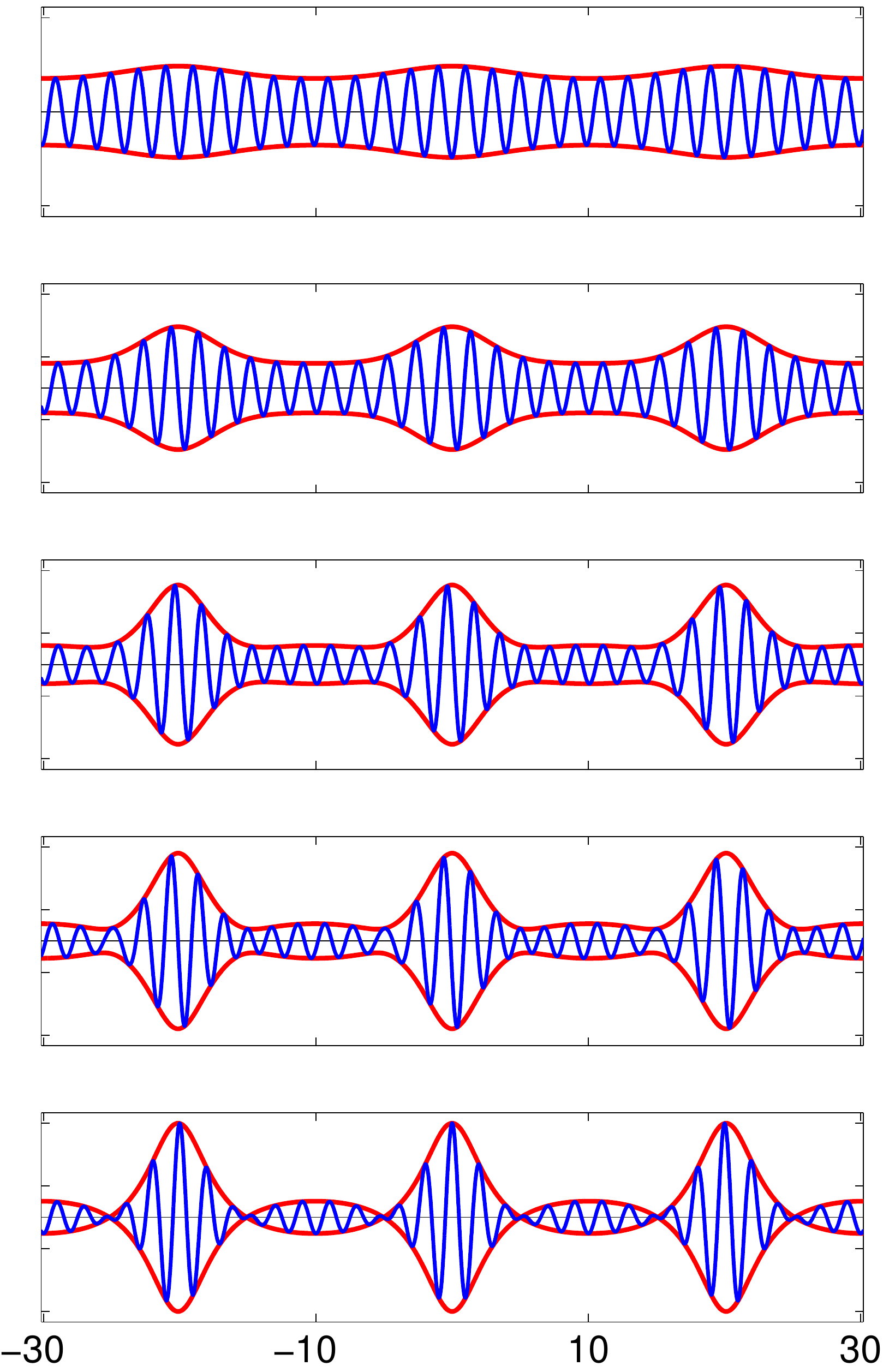,width = 0.5\textwidth}
    \else
      \epsfig{file = SFB1_evolution.eps,width = 0.5\textwidth}
    \fi
  \end{center}
  \caption{\footnotesize{The evolution of the SFB for $\tilde{\nu}_{1} = 1$
  from a modulated continuous wave signal into the extreme position.
  From top to bottom, the signals are taken at $x = -200$, $x = -100$,
  $x = -50$, $x = -30$, and $x = 0$.}}
  \label{SFB_evolution}
\end{figure}

By writing the complex-valued function $\psi$ in a polar (or
phase-amplitude) representation, it is found that for modulation
frequencie $\tilde{\nu}$ in the range
$0 < \tilde{\nu} < \sqrt{\frac{3}{2}}$ a \textit{phase singularity}
phenomenon occurs. It happens when the
real-valued amplitude $|\psi|$ vanishes and therefore there is no way of
ascribing a value to the real-valued phase when it occurs. The
local wave number $k\equiv k_0 + \partial_x \theta$ and
local frequency $\omega \equiv \omega_0 - \partial_t \theta$, with
$\theta(\xi,\tau)\equiv \arg{(\psi(\xi,\tau)}$,
become unbounded when this happens. The corresponding physical
wave packet profile $\eta(x,t)$ confirms this by showing a
\textit{wave dislocation} phenomenon. When the
real-valued amplitude $|\psi|$ vanishes at that specific position and time,
waves merge or split. For $\sqrt{\frac{3}{2}} < \tilde{\nu}
< \sqrt{2}$, the real-valued amplitude is always definite
positive, and thus there is no wave dislocation. Furthermore, in
one modulation period, there is a pair of wave dislocations.
Before or after this dislocation, the real-valued amplitude
reaches its maximum value. Figure \ref{SFBphysical} shows the
density plot of a physical wave packet profile $\eta(x,t)$. The wave
dislocation is also visible in this figure. Figure
\ref{SFB_evolution} shows the evolution of the SFB from a
modulated wave signal until it reaches the extreme position. We
can see also in this figure that the amplitude $|\psi|$ vanishes at
some moments for the extremal position $x=0$, causing phase singularity.

The phase singularity is a well known phenomenon in physical
optics. In the context of water waves, similar observations can be made,
and also wave dislocations occur. Trulsen \cite{Trul98} calls it
as \textit{crest pairing} and \textit{crest splitting} and he
explains this phenomenon as a consequence of linear dispersion.

\subsection*{Maximum temporal amplitude}
The maximum temporal amplitude (MTA) is a useful concept to understand
long-time behavior of wave elevation. For wave propagation in the
laboratory, it also
gives a direct view of the consequences of an initial wave signal on
the corresponding extreme-wave signal. It is defined as
\begin{equation}
  \mu(x) = \textmd{max}_{t}\; \eta(x,t),
\end{equation}
where $\eta(x,t)$ is the surface elevation as a function of space $x$
and time $t$. It describes the largest wave elevation that can appear
at a certain position. For laboratory wave generation, it describes the
boundary between the wet and dry parts of the wall of the basin after a
long time of wave evolution.

\begin{figure}[ht]
  \begin{center}
    \ifpdf
      \epsfig{file = 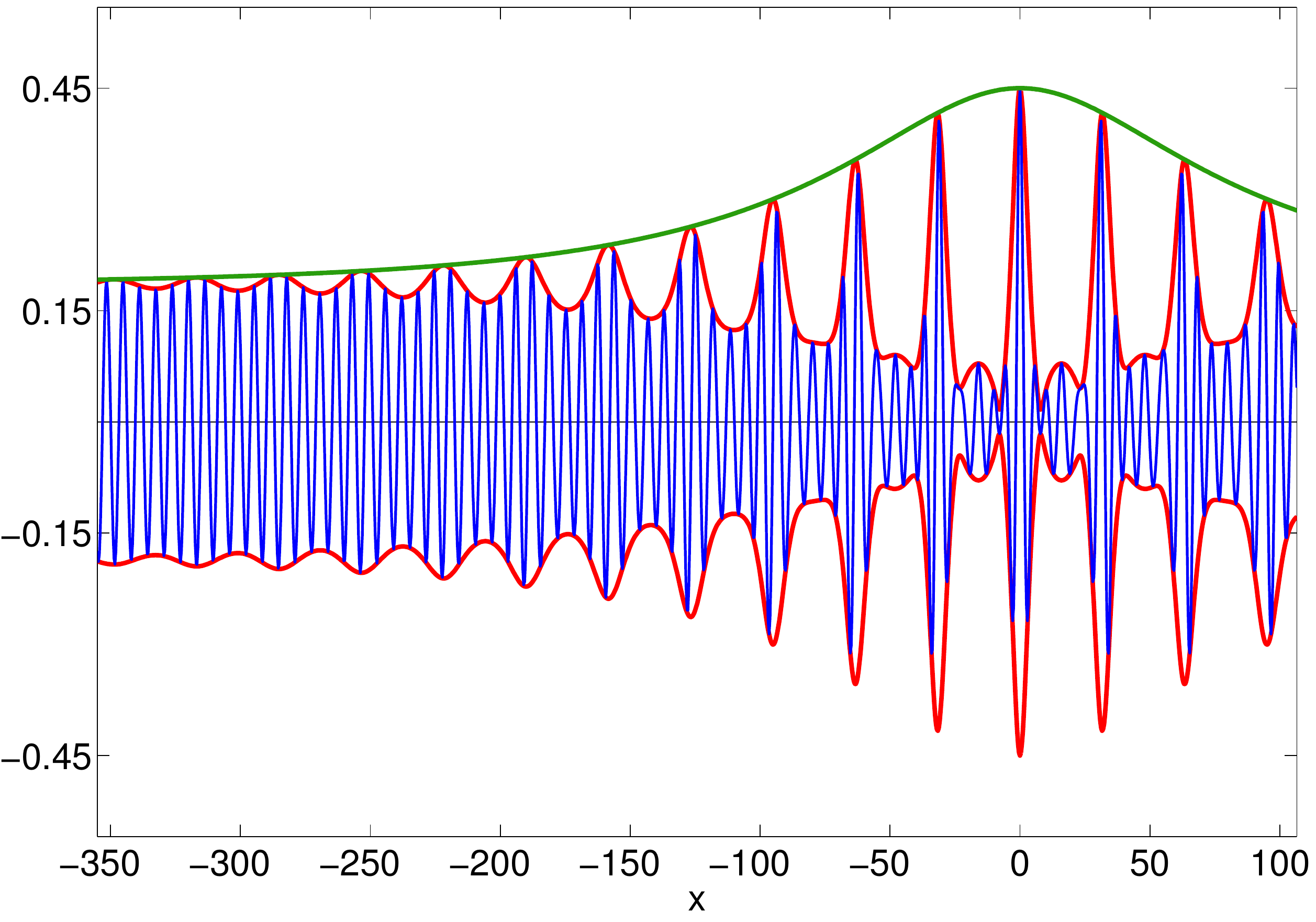,width = 12cm, height = 5cm, angle = 0}
    \else
      \epsfig{file = MTA1_profile.eps,width = 12cm, height = 5cm, angle = 0}
    \fi
  \end{center}
  \caption{\footnotesize{The MTA plot of the SFB, the corresponding wave profiles at $t = 0$, and its envelope.}}
  \label{MTA}
\end{figure}

Figure \ref{MTA} shows the MTA plot of the SFB in the laboratory
coordinates. In this example, the mean water depth is $3.55$ m and the
wavelength is approximately $6.2$ m. The wave signal is generated
at the left side, for example at $x_{\textmd{lab}} = -350$ m, and
it propagates to the right and reaches its extremal condition at
$x_{\textmd{lab}} = 0$. A slightly modulated wave train increases
in amplitude as the SFB waves travels in the positive $x$-direction.
Furthermore, in this example a SFB wave signal with initial
amplitude around $0.19$ m can reach an extreme amplitude of $0.45$
m, an amplification factor of around 2.4. After reaching its
maximum amplitude, the MTA decreases monotonically and returns to its
initial value.

\section*{Experimental Result}
\begin{figure}[ht]
  \begin{center}
    \ifpdf
      \epsfig{file = 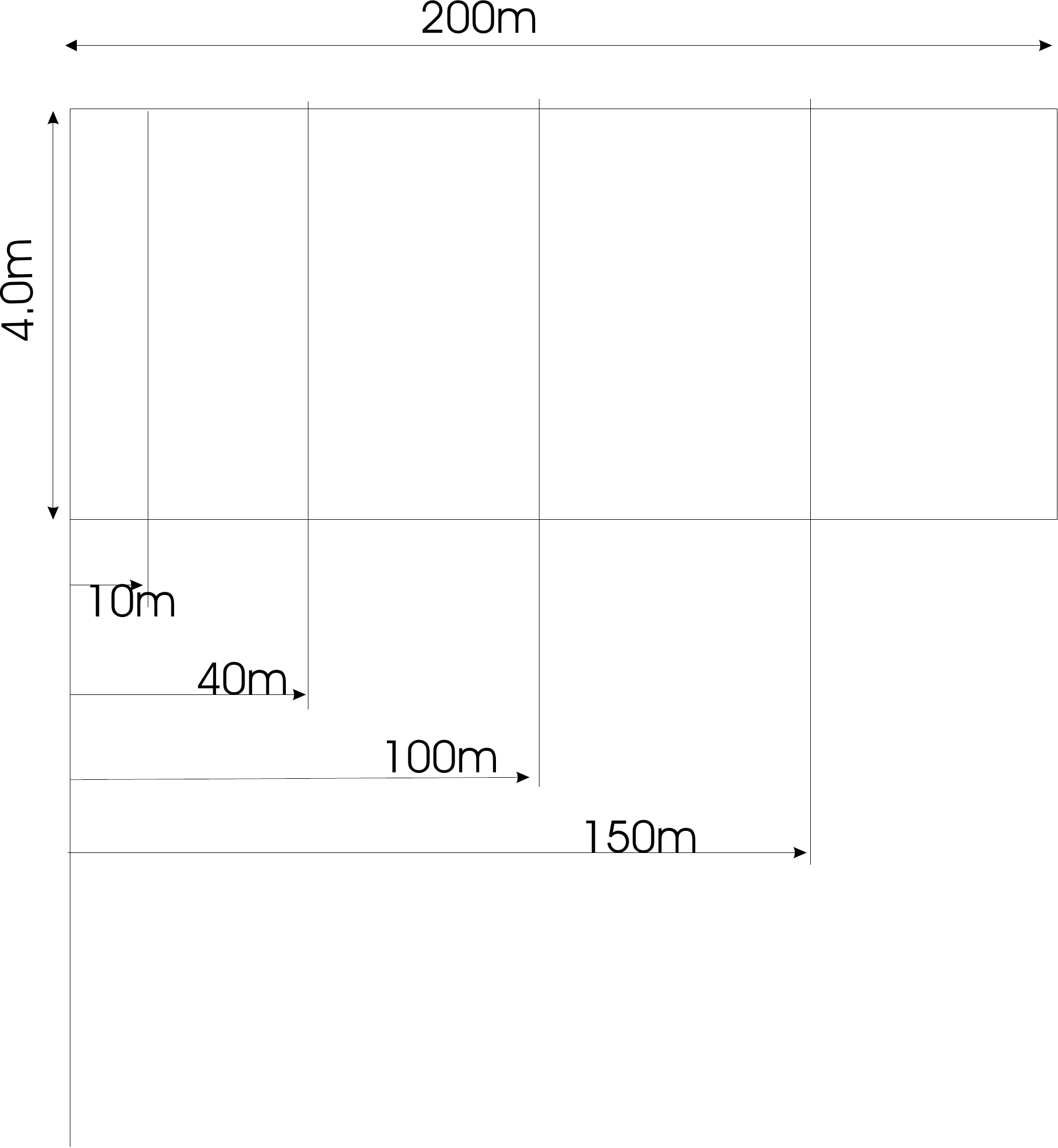, height= 5cm, width=10cm}
    \else
      \epsfig{file = set-up-ht.eps, height= 5cm, width=10cm}
    \fi
  \end{center}
  \caption{\footnotesize{The set-up of the wave probe array in the wave basin}}
  \label{set-up-ht}
\end{figure}

\begin{figure}[ht]
  \begin{center}
    \ifpdf
      \epsfig{file = 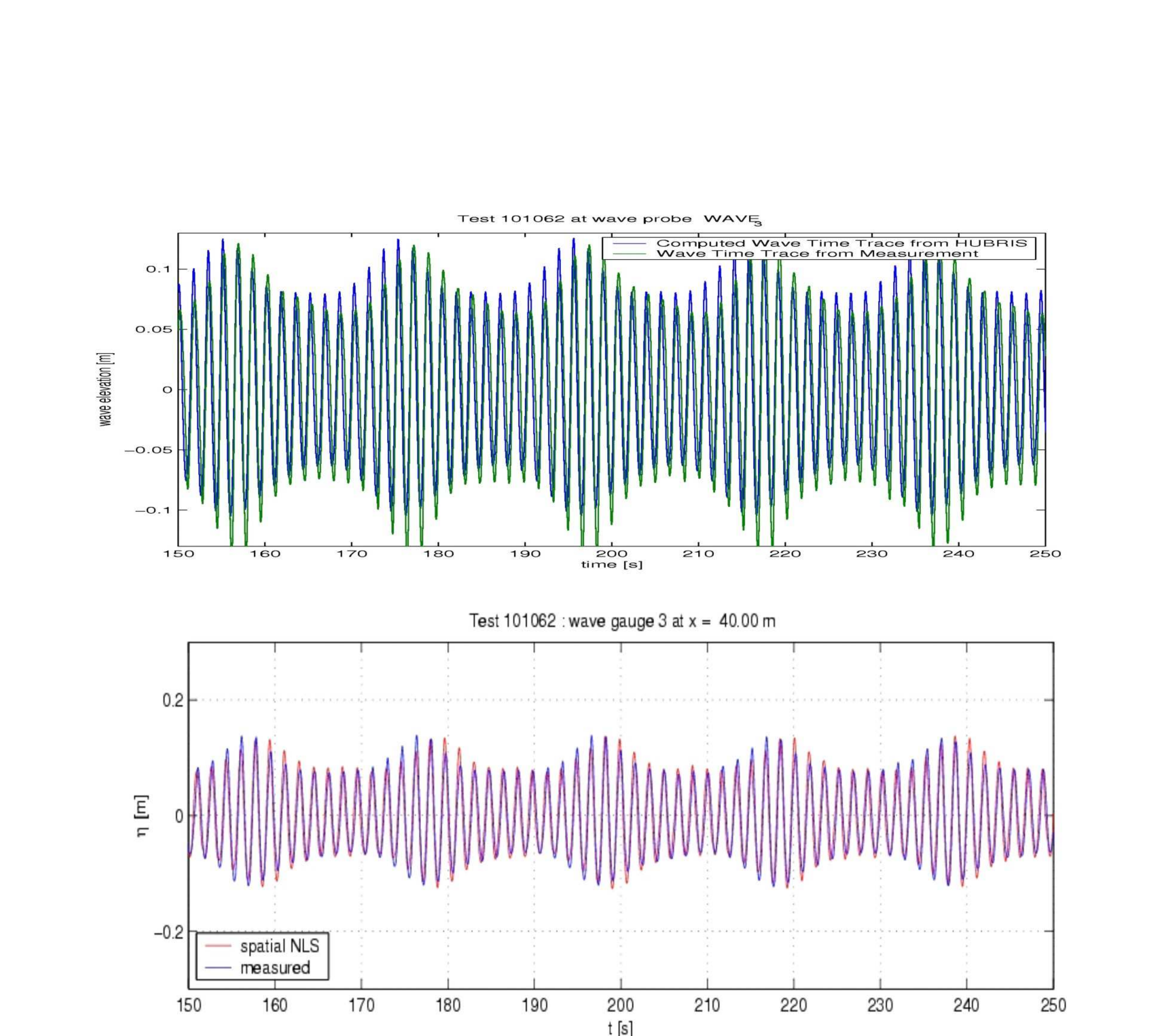, height= 8cm, width=12cm}
    \else
      \epsfig{file = hubris-snls.eps, height= 8cm, width=12cm}
    \fi
  \end{center}
  \caption{\footnotesize{Comparison Non-Linear Wave model HUBRIS with results from experiments and sNLS}}
  \label{hubris-snls}
\end{figure}

\begin{figure}[ht]
  \begin{center}
    \ifpdf
      \epsfig{file = 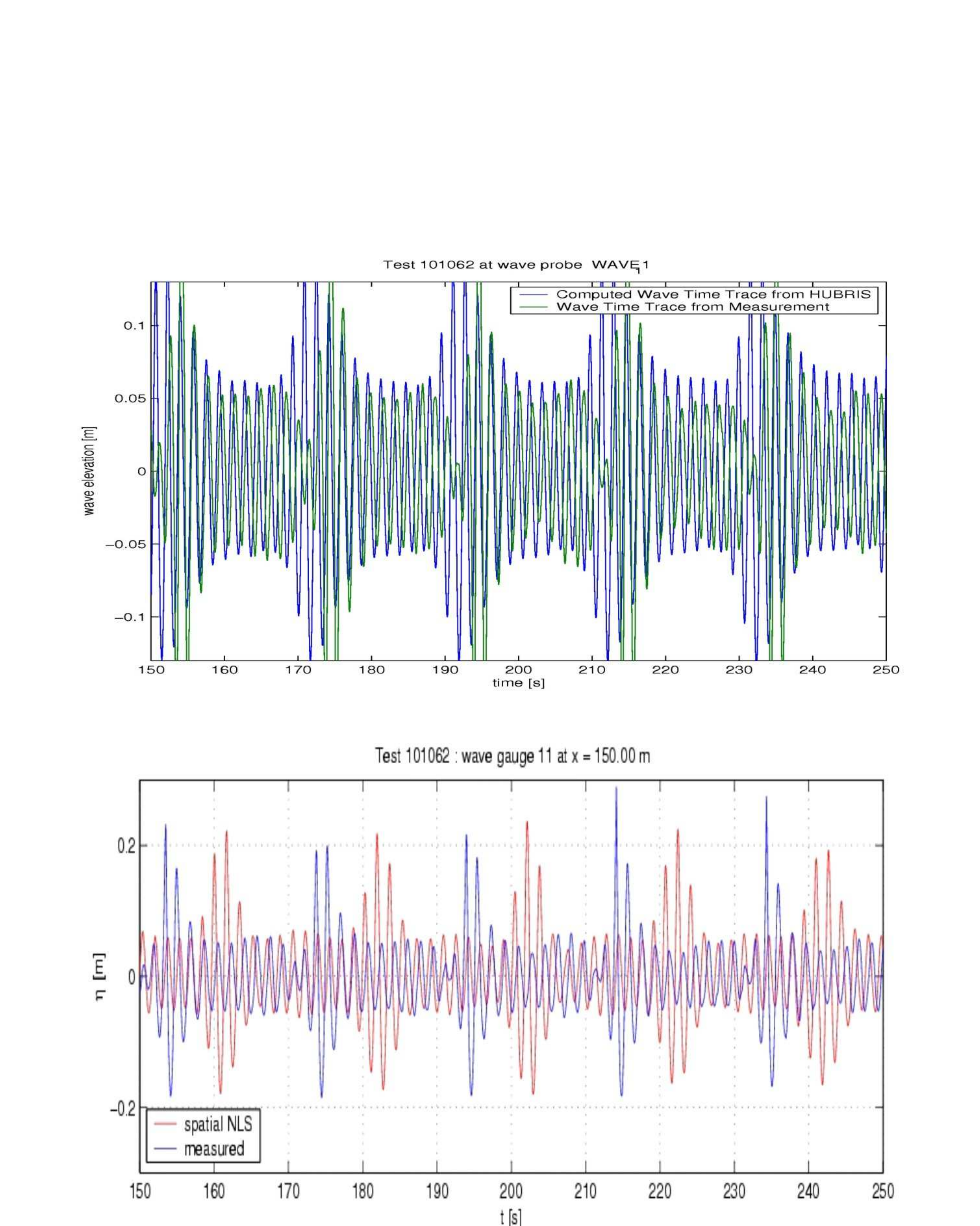, height= 8cm, width=12cm}
    \else
      \epsfig{file = hubris-snls3.eps, height= 8cm, width=12cm}
    \fi
  \end{center}
  \caption{\footnotesize{Comparison Non-Linear Wave model HUBRIS with results from experiments and sNLS}}
  \label{hubris-snls3}
\end{figure}

For the validation of the proposed method we performed experiments in one of the wave basins of \marin.
The basin dimensions amounted to $L \times B \times D$ as
200~m~$\times$~4.0~m~$\times$~3.55~m.
In the basin an array of wave probes were mounted as indicated in the set--up in figure (\ref{set-up-ht}).
The predefined wave board control signal was put onto the hydraulic wave generator. The stroke of the wave flap was measured.
Main characteristics of the model test experiments: \\
Carrier wave period is 1.685 sec. maximum wave height to be achieved (MTA) varies from 0.213m to 0.2485m\\
As a explanation the results for the tests with an MTA of 0.2485 will be shown,
 see figure( \ref{hubris-snls}) to figure(\ref{hubris-snls3}).
\begin{figure}[ht]
  \begin{center}
    \ifpdf
      \epsfig{file = 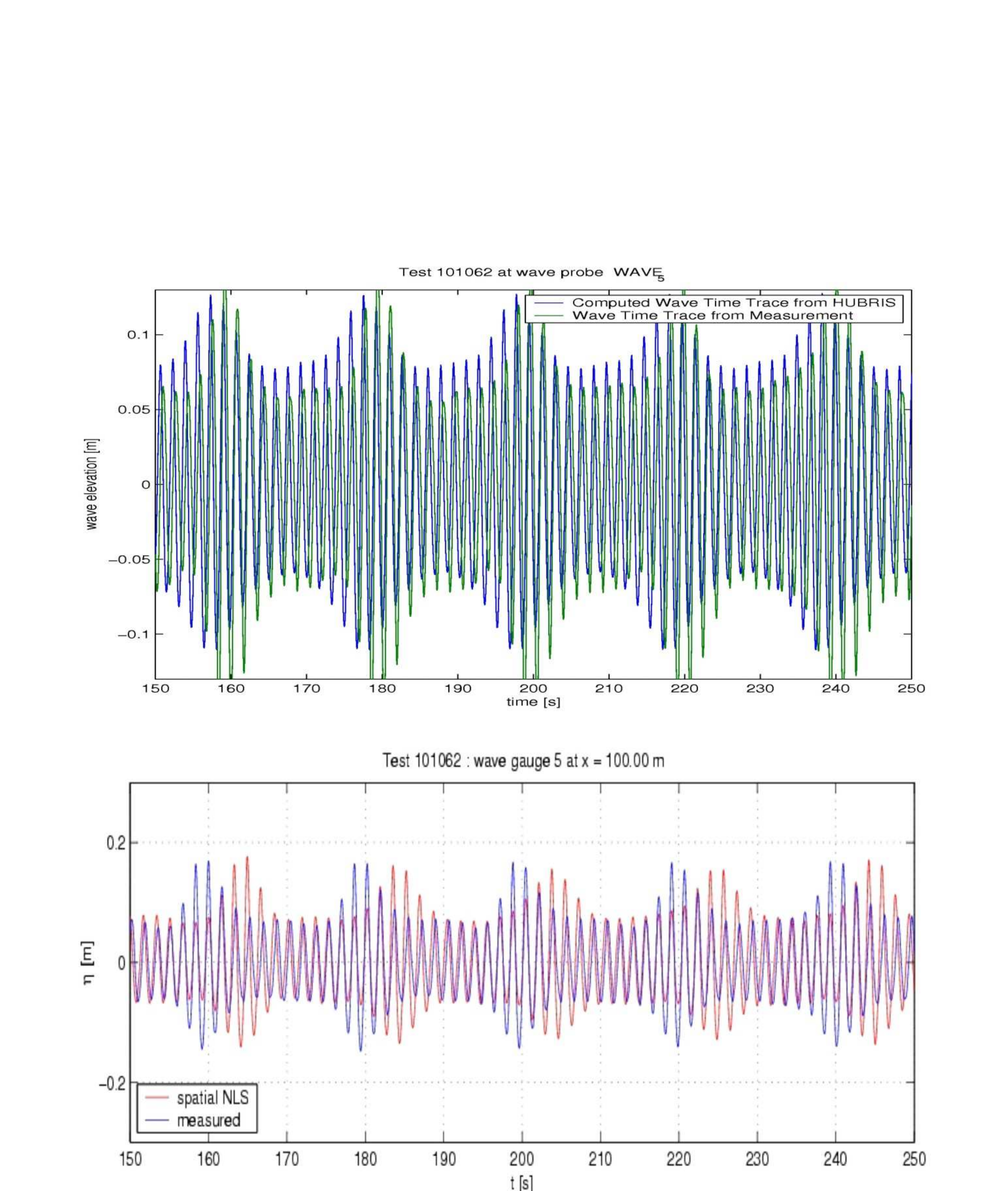, height= 8cm, width=12cm}
    \else
      \epsfig{file = hubris-snls2.eps, height= 8cm, width=12cm}
    \fi
  \end{center}
  \caption{\footnotesize{Comparison Non-Linear Wave model HUBRIS with results from experiments and sNLS}}
  \label{hubris-snls2}
\end{figure}

\begin{figure}[ht]
  \begin{center}
    \ifpdf
      \epsfig{file = 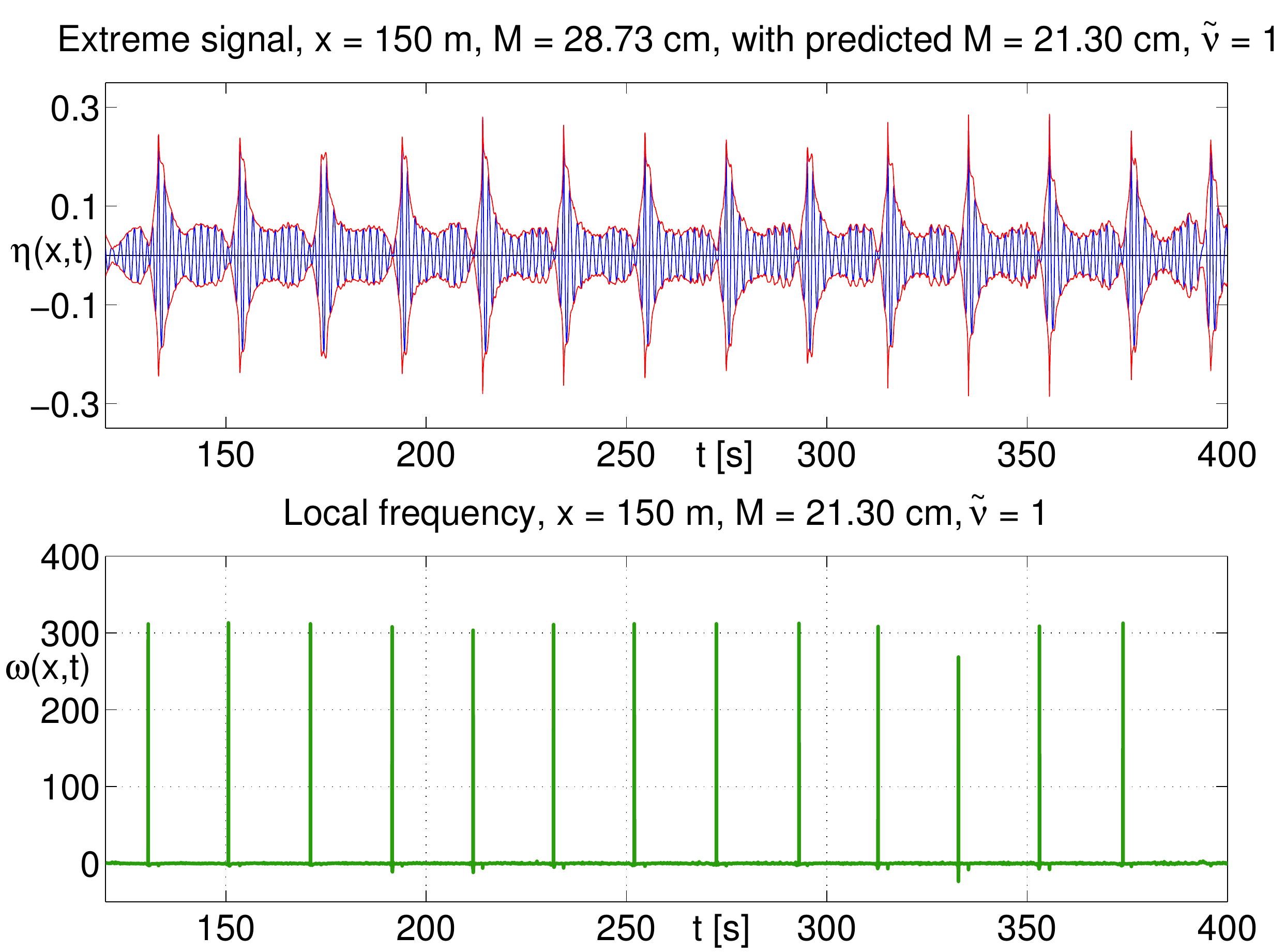,
              width=12.5cm, height=5cm, angle = 0}
    \else
      \epsfig{file = SFB1_locfreq_x150_M0213.eps,
              width=12.5cm, height=5cm, angle = 0}
    \fi
  \end{center}
  \caption{\footnotesize The SFB signal plot based on the experiment at 150 m (top) and
  the corresponding local frequency plot (bottom).}
  \label{signal_locfreq}
\end{figure}

Figure \ref{signal_locfreq} shows the SFB signal based on the
experiment at distance 150 m from the wave maker, where it is
expected that the signal to be extreme. That figure also shows the
phase singularity phenomenon when the local frequency becomes
unbounded when the real-valued amplitude vanishes or almost
vanishes. The experiment result shows asymmetric form of the
extreme signal while the theoretical result of the SFB preserve
the symmetry of the signal. It is suspected that if the modified
NLS equation of Dysthe \cite{Dyst79} is used as the governing
equation for the wave signal evolution, then there are good
comparisons with experimental measurements. The good comparisons
are observed for the case of bi-chromatic waves, where the
modified NLS equation predicts both the evolution of individual
wave crests and the modulation of the envelope over longer fetch
\cite{TrSt01}.
%\section*{Discussion and conclusions}

\end{document}